\documentclass[a4paper,12pt]{article}
\pdfoutput=1
\usepackage{feynmp-auto,expdlist}
\usepackage{amsmath, amsfonts, amssymb}
\usepackage{graphicx}
\usepackage{enumerate}
\usepackage{hyperref}
\usepackage{latexsym}
\usepackage{dsfont}
\usepackage{hepnicenames}
\usepackage{enumerate}
\usepackage{soul}
\usepackage[normalem]{ulem}
\usepackage{comment}

\usepackage{units}
\usepackage{mathrsfs,graphicx,rotating,amsmath,amsfonts,mathtools,booktabs,amssymb,wasysym}
\usepackage{hyperref}\usepackage{slashed}
\usepackage[nosort]{cite}
\usepackage[table,xcdraw,dvipsnames]{xcolor}
\usepackage{bm}
\usepackage{graphicx}
\usepackage{multirow,multicol}
\hypersetup{colorlinks,bookmarksopen,bookmarksnumbered,
	linkcolor=blus,pdfstartview=FitH,urlcolor=rossos,citecolor=verde}
\allowdisplaybreaks

\newcommand{\tauDG}{\tau_{\scriptscriptstyle{\rm DG}}}

\newcommand{\GammaDG}{\Gamma_{\scriptscriptstyle{\rm DG}}}

\renewcommand{\[}{\left[}

\def\Lag{\mathscr{L}}

\newcommand{\mio}[1]{}

\newcommand{\med}[1]{\langle #1\rangle}

\def\bpm{\begin{pmatrix}}
	\def\epm{\end{pmatrix}}

\usepackage{mathrsfs}

\newcommand{\fig}[1]{~\ref{fig:#1}}
\newcommand{\sfrac}[2]{#1/#2}

\allowdisplaybreaks
\usepackage{multicol}
\usepackage{color}
\definecolor{rosso}{cmyk}{0,1,1,0.4}
\definecolor{rossos}{cmyk}{0,1,1,0.55}
\definecolor{rossoc}{cmyk}{0,1,1,0.2}
\definecolor{blu}{cmyk}{1,1,0,0.3}
\definecolor{blus}{cmyk}{1,1,0,0.6}
\definecolor{bluc}{cmyk}{1,1,0,0.1}
\definecolor{verde}{cmyk}{0.92,0,0.59,0.25}
\definecolor{verdec}{cmyk}{0.92,0,0.59,0.15}
\definecolor{verdes}{cmyk}{0.92,0,0.59,0.4}

\newcommand{\bp}{\bar{M}_{\rm Pl}}
\newcommand\Ord{{\cal O}}
\newcommand{\eq}[1]{~{\rm (\ref{eq:#1})}}

\newcommand{\s}{\,{\rm s}}

\newcommand{\GeV}{\,{\rm GeV}}
\newcommand{\TeV}{\,{\rm TeV}}
\newcommand{\cm}{\,{\rm cm}}

\newcommand{\Tr}{\,{\rm Tr}}

\def\circa#1{\,\raise.3ex\hbox{$#1$\kern-.75em\lower1ex\hbox{$\sim$}}\,}

\newcommand{\beq}{\begin{equation}}
\newcommand{\eeq}{\end{equation}}

\newcommand{\bea}{\begin{eqnarray}}
\newcommand{\eea}{\end{eqnarray}}
\newcommand{\be}{\begin{equation}}
\newcommand{\ee}{\end{equation}}
\font\tenrsfs=rsfs10 at 12pt
\font\sevenrsfs=rsfs7
\font\fiversfs=rsfs5
\newfam\rsfsfam
\textfont\rsfsfam=\tenrsfs
\scriptfont\rsfsfam=\sevenrsfs
\scriptscriptfont\rsfsfam=\fiversfs

\newcommand{\bra}{\langle}
\newcommand{\ket}{\rangle}

\newsavebox\MBox

\newcommand{\Sp}{\,{\rm Sp}}

\renewenvironment{thebibliography}[1]
{\begin{multicols}{2}[\section*{\refname}]%
		\@mkboth{\MakeUppercase\refname}{\MakeUppercase\refname}%
		\list{\@biblabel{\@arabic\c@enumiv}}%
		{\settowidth\labelwidth{\@biblabel{#1}}%
			\leftmargin\labelwidth
			\advance\leftmargin\labelsep
			\@openbib@code
			\usecounter{enumiv}%
			\let\p@enumiv\@empty
			\renewcommand\theenumiv{\@arabic\c@enumiv}}%
		\sloppy
		\clubpenalty4000
		\@clubpenalty \clubpenalty
		\widowpenalty4000%
		\sfcode`\.\@m}
	{\def\@noitemerr
		{\@latex@warning{Empty `thebibliography' environment}}%
		\endlist\end{multicols}}

\newcommand{\SU}{\,{\rm SU}}
\newcommand{\SO}{\,{\rm SO}}

\def\circa#1{\,\raise.3ex\hbox{$#1$\kern-.75em\lower1ex\hbox{$\sim$}}\,}
\makeatletter

\font\ital=cmu10

\def\hhref#1{\href{http://arxiv.org/abs/#1}{arXiv:#1}}
\usepackage{xstring}
\newcommand{\hhrefq}[1]{\IfSubStr{#1}{:}{\href{http://inspirehep.net/search?ln=en&ln=en&p=#1&of=hb&action_search=Search&sf=&so=d&rm=&rg=25&sc=0}{InSpire:#1}}{\hhref{#1}}}

\def\art{\@ifnextchar[{\eart}{\oart}}
\def\eart[#1]#2#3#4#5#6{{\rm #2}, {\em #3 \bf #4} {\rm (#6) #5} ({\em #1})}
\def\article{\@ifnextchar[{\earticle}{\oarticle}}
\def\oarticle#1#2#3#4#5#6{{\rm #1}, {\ital ``#6''}, {\rm #2 #3 (#5) #4}}
\def\earticle[#1]#2#3#4#5#6#7{{\rm #2}, {\ital ``#7''}, {\rm #3 #4 (#6) #5}  [\hhrefq{#1}]}
\def\hepart[#1]#2{{\rm #2, \sl#1}}
\def\heparticle[#1]#2#3{#2, {\ital ``#3''} [\hhrefq{#1}]}
\newcommand{\doi}[1]{\href{http://dx.doi.org/#1}{[link]}}

\newcommand{\hhrefqq}[1]{\IfBeginWith{#1}{10.}{\href{https://doi.org/#1}{doi:#1}}{\hhrefq{#1}}}
\def\earticle[#1]#2#3#4#5#6#7{{\rm #2}, {\ital ``#7''}, {\rm #3 #4 (#6) #5}  [\hhrefqq{#1}]}
\renewenvironment{thebibliography}[1]
{\begin{multicols}{2}[\section*{\refname}]%
		\@mkboth{\MakeUppercase\refname}{\MakeUppercase\refname}%
		\list{\@biblabel{\@arabic\c@enumiv}}%
		{\settowidth\labelwidth{\@biblabel{#1}}%
			\leftmargin\labelwidth
			\advance\leftmargin\labelsep
			\@openbib@code
			\usecounter{enumiv}%
			\let\p@enumiv\@empty
			\renewcommand\theenumiv{\@arabic\c@enumiv}}%
		\sloppy
		\clubpenalty4000
		\@clubpenalty \clubpenalty
		\widowpenalty4000%
		\sfcode`\.\@m}
	{\def\@noitemerr
		{\@latex@warning{Empty `thebibliography' environment}}%
		\endlist\end{multicols}}

\newcounter{alphaequation}[equation]
\def\thealphaequation{\theequation\hbox to
	0.6em{\hfil\alph{alphaequation}\hfil}}
\def\eqnsystem#1{
	\def\@eqnnum{{\rm (\thealphaequation)}}
	\def\@@eqncr{\let\@tempa\relax \ifcase\@eqcnt \def\@tempa{& & &} \or
		\def\@tempa{& &}\or \def\@tempa{&}\fi\@tempa
		\if@eqnsw\@eqnnum\refstepcounter{alphaequation}\fi
		\global\@eqnswtrue\global\@eqcnt=0\cr}
	\refstepcounter{equation} \let\@currentlabel\theequation \def\@tempb{#1}
	\ifx\@tempb\empty\else\label{#1}\fi
	\refstepcounter{alphaequation}
	\let\@currentlabel\thealphaequation
	\global\@eqnswtrue\global\@eqcnt=0 \tabskip\@centering\let\\=\@eqncr
	$$\halign to \displaywidth\bgroup \@eqnsel\hskip\@centering
	$\displaystyle\tabskip\z@{##}$&\global\@eqcnt\@ne
	\hskip2\arraycolsep\hfil${##}$\hfil& \global\@eqcnt\tw@\hskip2\arraycolsep
	$\displaystyle\tabskip\z@{##}$\hfil
	\tabskip\@centering&\llap{##}\tabskip\z@\cr}
\def\endeqnsystem{\@@eqncr\egroup$$\global\@ignoretrue} \makeatother

\newcommand{\GL}[1]{{{[\bf GL: #1]}}}

\definecolor{Gray}{gray}{0.95}

\def\bal#1\eal{\begin{align}#1\end{align}}

\newcommand{\N}{\mathcal{N}}
\newcommand{\G}{\mathcal{G}}
\newcommand{\gdc}{g_{\scriptstyle{\rm DM}}}
\newcommand{\ldc}{\Lambda_{\scriptstyle{\rm DM}}}
\newcommand{\adc}{\alpha_{\scriptstyle{\rm DM}}}
\newcommand{\mdg}{M_{\scriptstyle{\rm DG}}}
\newcommand{\trh}{T_{\rm RH}}
\newcommand{\hi}{H_{\rm infl}}

\begin{document}
\thispagestyle{empty}

\begin{center}  %\vspace{3cm}
{\Large\Huge\bf\color{rossos} Gravitational Vector Dark Matter} \\
\vspace{1cm}
{\bf 
Christian Gross$^{a,b}$,
Sotirios Karamitsos$^{a}$,
\\
Giacomo Landini$^{a,b}$, 
Alessandro Strumia$^{a}$}\\[7mm]

{\it $^a$ Dipartimento di Fisica, Universit\`a di Pisa, Italy}\\[1mm]
{\it $^b$ INFN, Sezione di Pisa, Italy}\\[1mm]

\vspace{1cm}

{\large\bf Abstract}

\begin{quote}\large
A new dark sector consisting of a pure non-abelian gauge theory has no renormalizable interaction with SM particles, and can thereby realise gravitational Dark Matter (DM). Gauge interactions confine at a scale $\Lambda_{\rm DM}$ giving bound states with typical lifetimes $\tau \sim M_{\rm Pl}^4/\Lambda^5_{\rm DM}$ that can be DM candidates if $\Lambda_{\rm DM} $ is below 100 TeV. Furthermore, accidental symmetries of group-theoretical nature produce special gravitationally stable bound states. In the presence of generic Planck-suppressed operators such states become long-lived: SU$(N)$ gauge theories contain bound states with $\tau \sim M_{\rm Pl}^8/\Lambda^9_{\rm DM}$; even longer lifetimes $\tau=  (M_{\rm Pl}/\Lambda_{\rm DM})^{2N-4}/\Lambda_{\rm DM}$  arise from SO$(N)$ theories with $N \ge 8$, and possibly from $F_4$ or $E_8$. We compute their relic abundance generated by gravitational freeze-in and by inflationary fluctuations, finding that they can be viable DM candidates for $\Lambda_{\rm DM} \gtrsim 10^{10}$ GeV.
\end{quote}
\end{center}

\clearpage
\setcounter{page}{1}

\tableofcontents
	
\section{Introduction}
A new quasi-stable particle with mass $M$, spin 0, 1/2 or 1 and gravitational interactions only
is a phenomenologically viable DM candidate, dubbed `gravitational DM'~\cite{Garny:2015sjg,1708.05138,1709.09688}.
Such DM can be produced through gravitational scatterings
or through  fluctuations during inflation.

Can gravitational DM be realised  in reasonable theories, 
or does it need ad hoc assumptions?
For example a scalar $S$ can always have non-gravitational renormalizable quartic interactions
to the Higgs $H$, $|H|^2|S|^2$.
A singlet fermion $N$ can always have a renormalizable Yukawa interaction to
the Higgs and left-handed leptons $L$, $NLH$.
{(This can be forbidden imposing a $\mathbb{Z}_2$ $N\to -N$ symmetry)}.
The  models of vectors considered in~\cite{Garny:2015sjg,1708.05138,1709.09688} 
have the same problem, as they involve a scalar $S$ to make vectors massive;
furthermore abelian vectors can mix with hypercharge at renormalizable level.
We thereby here consider a theory based on the SM plus a new non-abelian gauge group $G$
and no scalars or fermions charged under it.
The action is
\beq S =  \int d^4x \sqrt{|\det g|} \left[- \frac12 \bar M_{\rm Pl}^2 R + \Lag_{\rm SM } +\Lag_{\rm DM}  +
\Lag_{\rm NRO} \right]
\eeq
where $M_{\rm Pl}=\sqrt{8\pi}\bar M_{\rm Pl} =1.2\times10^{19}$ GeV is the Planck mass,
and $\Lag_{\rm SM}$ and $\Lag_{\rm DM}$ describe the renormalizable interactions in the SM and DM sectors.
As the dark sector is a pure gauge theory with non-abelian gauge group $G$, at renormalizable level  DM is decoupled from the SM. 
The
most generic Lagrangian is
\beq\label{eq:LDM}
\Lag_{\rm DM} = - \frac14 {G}_{\mu\nu}^a G^{\mu\nu a} + 
\theta_{\rm DM} \frac{\gdc^2}{32\pi^2}  {G}_{\mu\nu}^a \tilde{G}^{\mu\nu a}\eeq
where $G_{\mu\nu}^a = \partial_\mu G_\nu^a- \partial_\nu G_\mu^a- \gdc f^{abc} G^b_\mu G^c_\nu$.
The dark $\theta_{\rm DM}$ term is physical and non-perturbatively breaks P and CP in the dark sector;
we assume that $\theta_{\rm DM}$ is of order unity, and it will not give qualitatively new effects.
Given that the dark sector confines at a scale $\ldc$,  the glueball hadrons made of vectors are possible
DM candidates provided that they are long lived enough. 
Different decay rates arise depending on the possible reasonable assumptions considered below.
\begin{enumerate}
\item In the most extreme case one could assume that
$\Lag_{\rm NRO}=0$ i.e.\ that gravity is the only non-renormalizable interaction.
This is a consistent possibility in theories of renormalizable quantum gravity based on 4-derivatives,
that however contain possibly problematic states with negative classical energy (see e.g.~\cite{Stelle:1976gc,1403.4226}).
Under this assumption, most dark bound states undergo gravitational decays with rates $\Gamma \sim m^5/M_{\rm Pl}^4$.
They can be DM only if long-lived, which needs their mass $m$ to be below $\sim 100\TeV$.
We will find that some gauge groups $G$ predict other bound states that are exactly stable, in this limit.

\item In more general theories,  gravity and SM interactions give rise,
via perturbative quantum corrections such as renormalisation group (RG) running,
to Planck-suppressed non-renormalizable operators 
that link the SM and DM sectors and respect the accidental symmetries of the renormalizable Lagrangian $\Lag_{\rm SM}+\Lag_{\rm DM}$.
One example is dimension 6 operators such as $\Lag_{\rm DM}|H|^2/M_{\rm Pl}^2$~\cite{1602.00714}.
Such operators correct the rates of gravitational processes by order unity factors.

\item As an alternative possibility, generic Planck-suppressed non-renormalizable operators might be present in 
some theories of quantum gravity, such as those with lots of new states around the Planck scale (e.g.\ string models),
and those where gravity gets strongly coupled (non-perturbative quantum gravity, possibly dominated by
black holes and wormholes, is expected to violate accidental symmetries~\cite{Abbott:1989jw,1807.00824}).
Such operators imply decays of generic dark bound states.
Some bound states remain long-lived enough to be DM candidates, even if heavy.
\end{enumerate}
The paper is structured as follows.

In section~\ref{glueballs} we study the bound states and their lifetimes, finding,
in addition to the ordinary glueballs, 
special longer-lived glueballs if
the gauge group is $G=\SU(N)$ and very long-lived states  if
the gauge group $\SO(2N)$ at large $N\circa{>}8$.
Such states can be DM candidates even if heavier than about $10^{10}\GeV$.

In section~\ref{thermal} we compute DM production via gravitational thermal freeze in.
This is maximally efficient if the reheating temperature $T_{\rm RH}$ is comparable to $\ldc$.
Our results differ from~\cite{Garny:2015sjg,1708.05138,1709.09688} that considered gravitational production of massive vectors with 3 degrees of freedom, as our  massless vectors have 2 degrees of freedom, and acquire mass through confinement.
Our study is similar to \cite{2011.10565}, that however only considered gravitational freeze-in of $\SU(N)$ ordinary glueballs.\footnote{Other works considered confined vectors in the opposite limit, where $\SU(N)$ glueballs interact so strongly via enhanced NRO with the SM sector, that 
the thermal relic abundance is relevant~\cite{1602.00714,1605.08048,1710.06447}.}

In section~\ref{inflation} we consider gravitational production during inflation.
As gravitational production via scatterings is dominated by the highest reheating temperature after inflation, 
inflation itself can contribute more. 
Ignoring possibly large but model-dependent effects
(production of DM from inflaton decay or from post-inflationary inflaton oscillations, if kinematically allowed)
we find that pure gravitational production during inflation is sub-leading.
Indeed our (purely transverse) vectors are conformally coupled at tree level, so they are not produced in conformally flat cosmological backgrounds.
The conformal symmetry is broken by the quantum running of the gauge coupling $\gdc$ and by confinement, 
giving rise to computable effects.

Conclusions are given in section~\ref{concl}.

\section{Bound states of vectors}\label{glueballs}
Dark vectors are stable, being the only states charged under the dark gauge group.
However, at dark temperatures $T_{\rm DM}\circa{<}\ldc$  they acquire mass through confinement, forming 
dark glueballs (DG) with mass  $\mdg  \sim \ldc$. 
Thereby DM candidates arise if some bound state made of dark vectors is long-lived enough.
RG running of $\adc=\gdc^2/4\pi$  is given, in one loop approximation, by
\begin{equation}
\frac{1}{\adc(\mu)} = \frac{1}{\adc(\mu')} + \frac{b}{2\pi}\ln \frac{\mu}{\mu'},\qquad b =  \frac{11}{3} C_G
\end{equation}
where $C_G$ is the quadratic Casimir of the group $G$.
Furthermore we define $d_G$ as the dimension of the group $G$, so that
$C_G = N$ and $d_G = N^2-1$ for $G=\SU(N)$; and $C_G = 2(N-2)$, $d_G = N(N-1)/2$  for $\SO(N)$.

So the running dark gauge coupling is related to the energy scale $\ldc$ at which the dark sector confines,
$\adc(\ldc)\sim 4\pi$, by
\begin{equation}\label{eq:RGEsol}
\adc(\mu) \approx \frac{2\pi}{b} \frac{1}{\ln\mu/\ldc}.
\end{equation}

\begin{figure}[t]
\centering
$$\includegraphics[width=0.95\textwidth]{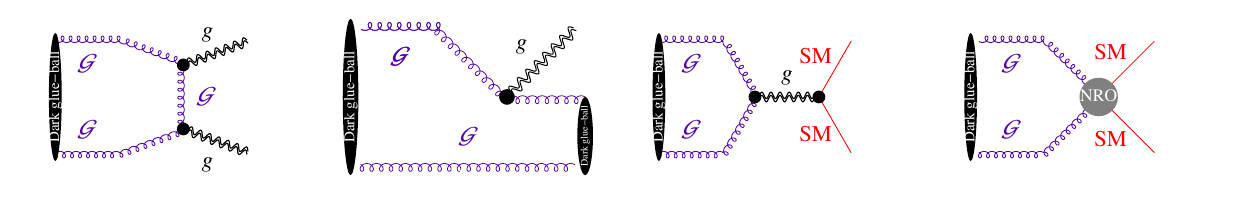}$$
\vspace{-1cm}
\caption{\em\label{fig:DGgravitationalDecays}
Possible gravitational decays of dark glueballs. SM denotes any Standard Model particle, including gravitons.}
\end{figure}

\subsection{Ordinary glueballs}
Analogy with QCD computations~\cite{Jaffe:1985qp,0903.0883,1301.5183} suggests that
the lightest dark glueball is the state with spin 0 and quantum numbers $J^{PC} = 0^{++}$
that corresponds to the gauge-invariant operator $\Tr \,[{\cal G}_{\mu\nu}{\cal G}^{\mu\nu}]$,
where $\G^i_j \equiv G^a (T^a)^i_j$ and $T^a$ are the generators in the fundamental.
We assume $\mdg  \approx 2 \ldc$.
Such state decays gravitationally into two gravitons, or into 
SM particles via one-graviton exchange (see fig.\fig{DGgravitationalDecays}).
Since two gravitational couplings are involved,
the decay amplitude arises at order $1/M_{\rm Pl}^2$, and thereby gives a decay width of order
\beq \label{eq:tauDG4}
\GammaDG \equiv 1/\tauDG \sim \ldc^5/M_{\rm Pl}^4 .\eeq
Order one couplings depend on non-perturbative factors.
Furthermore, generic Planck-suppressed operators would contribute giving comparable decay rates.
We thereby do not go beyond the estimate.

\begin{figure}[t]
\centering
$$\includegraphics[width=0.4\textwidth]{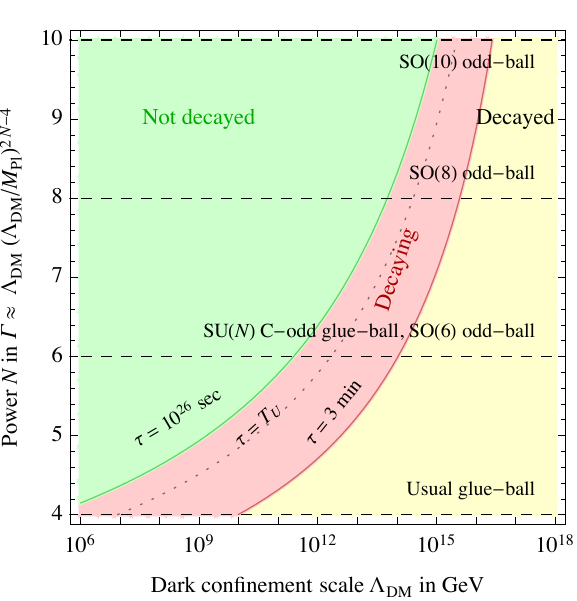}$$
\vspace{-1cm}
\caption{\em\label{fig:glueballLifeTime} Estimated lifetime of glueballs, compared to time-scales relevant for
indirect detection bounds and Big Bang Nucleosynthesis.}
\end{figure}

Particles with DM-like relic abundance that decay 
into SM states with lifetime $\tau$ in the range $ 3\,{\rm min}\circa{<}  \tau \circa{<} 10^{26}\s$ are excluded.
The lower bound is obtained from BBN, the upper bound from indirect detection
(see e.g.~\cite{0811.4153}) and the intermediate range is covered by CMB observations (see e.g.~\cite{Wright:1993re,1610.10051,1810.05912}).
Thereby, dark glueball DM with lifetime given by eq.\eq{tauDG4} is excluded
in the mass range $100\TeV \circa{<} \mdg \circa{<} 10^{10}\GeV$,
see fig.\fig{glueballLifeTime}.

\medskip

QCD computations also indicate that various heavier glueballs (pseudo-scalars, 
resonances with spin 2, as well as $\G\G\G$ states with spin 1 and possibly 3),
are light enough that they cannot undergo fast gauge decays into two lightest glueballs.
Most of such states (the exceptions will be discussed in the next sections)
undergo gravitational decays with rates comparable to $\GammaDG$,
possibly including decays that involve one lighter glueball
and that can be faster, $\Gamma({\rm DG}'\to {\rm DG} g)\sim \ldc^3/M_{\rm Pl}^2$,
see e.g.\ fig.\fig{DGgravitationalDecays}b.

The dark gauge group in general has a topological $\theta_{\textrm{DM}}$ term, that violates
space-time parity P and CP at non-perturbative level, such that P-even glueballs
(such as $\Tr\,\G\G$) mix with P-odd glueballs (such as $\Tr\, \G\tilde\G$).
The $\Tr\, \G\tilde\G$ glueball cannot decay gravitationally, but can decay via NRO with similar rate.

We next discuss the special glueballs, long-lived because of group-theoretical accidental symmetries.

\subsection{Long-lived SU glueballs: group charge conjugation}
The U(1), $\SU(N)$, $\SO(2k)$ and $E_6$ groups with symmetric Dynkin diagrams 
have complex representations,
and complex conjugation is a $\mathbb{Z}_2$ outer automorphism of the group~\cite{Wright:1993re,1608.05240,1610.10051,1810.05912}.
The explicit action of C can be obtained considering 
any complex representation with generators $T^a$:
gauge interactions are left invariant by
$\G\stackrel{\rm C}{\longrightarrow} - \G^*$.
So C-even vectors are associated to purely imaginary generators $T^a$.
The overall minus sign arises because, in field theory, vectors
couple to C-odd currents
(of scalars, $J_\mu\sim  \phi_* T^a \partial_\mu \phi$, of vectors and of fermions).
A U(1) vector is C-odd, $G_\mu \stackrel{\rm C}{\longrightarrow} - G_\mu$.
In a basis where each vector $G^a_\mu$ 
is either $O$dd or $E$ven under C the group structure constants $f_{abc}$ and the symmetric tensor $d_{abc}$ satisfy
$f_{OOO},f_{OEE}=0$, $f_{OOE}, f_{EEE}\neq 0$
and
$d_{OOO},d_{OEE}\neq 0$, $d_{OOE}, d_{EEE}= 0$.
So $G^a_{\mu\nu}$ has the same parity as $G^a_\mu$ and~\cite{1710.06447}
$$\begin{array}{rcll}
\Tr\,\G_{\mu\mu'} \G_{\nu\nu'} & \propto & G^a_{\mu\mu'} G^a_{\nu\nu'} & \hbox{is CP-even}\\
\Tr\, \G_{\mu\mu'}[ \G_{\nu\nu'} ,\G_{\rho\rho'}] & \propto & i f^{abc} G^a_{\mu\mu'} G^b_{\nu\nu'} G^c_{\rho\rho'}& \hbox{is CP-even}\\
\Tr\, \G_{\mu\mu'}\{\G_{\nu\nu'} ,\G_{\rho\rho'}\} & \propto &
d^{abc} G^a_{\mu\mu'} G^b_{\nu\nu'} G^c_{\rho\rho'}& \hbox{is CP-odd.}
\end{array}
$$
The $d$ tensor (related to anomalies, in theories with fermions)
is non-vanishing only for $\SU(N)$ groups, excluding $\SU(2)=\SO(3)$ and including $\SU(4)=\SO(6)$.
For all other groups the C-odd $\G\G\G$ glueball vanishes.
$\SO(2N)$ and $E_6$ groups admit complex representations, but $d_{abc}$ vanishes.

The C-odd glueball predicted by pure $\SU(N)$ groups\footnote{The C-odd QCD glueball
is unstable because of the presence of quarks and has been recently observed~\cite{2012.03981}.}
 is stable under gravitational decays,
because gravitational interactions respect C, so charge conjugation acting on
SM particles and on dark vectors are independent symmetries.
On the other hand, the C-odd glueball can decay gravitationally if the action contains
the non-renormalizable operator $\Tr(\G\{\G,\G\})/M_{\rm Pl}^2$, which explicitly breaks C-parity.
The C-odd state can decay into SM particles in the presence of dimension-8
operators such as
$\Tr(\G^3)|H|^2/M_{\rm Pl}^4$ or $\Tr(\G^3)_{\mu\nu}B^{\mu\nu}/M_{\rm Pl}^4$.
In both cases, the C-odd ball acquires a decay rate 
\beq \label{eq:GammaCB}
\Gamma_{\rm C-odd} \sim \ldc^9/M_{\rm Pl}^8.\eeq
C-odd $\SU(N)$ glueballs could be the only DM in a narrow mass range around $\ldc \sim  10^{11}\GeV$, see fig.\fig{glueballLifeTime}.

\begin{table}[t]
\begin{center}
$$\begin{array}{c|ccc}
\hbox{group} & \hbox{fundamental} & \hbox{adjoint} & \hbox{invariant tensors} \\ \hline
\SU(N) & N, \bar N & N^2-1 & \delta_i^j, \epsilon_{i_1\cdots i_N}\\
\SO(N) & N & N(N-1)/2 & \delta_{ij} , \epsilon_{i_1\cdots i_N}\\
\Sp(2N) & 2N & N(2N+1) &A_{ij}, \epsilon_{i_1\cdots i_{2N}}, \\
G_2 & 7 & 14 &\delta_{ij}, A_{ijk}, \epsilon_{i_1\cdots i_7}\\
F_4 \subset \SO(26)& 26 & 52 & \delta_{ij}, S_{ijk}, \epsilon_{i_1\cdots i_{26}}\\
E_6 \subset\SU(27) &27, \overline{27} & 78 & \delta_{i}^j, S_{ijk}, \epsilon_{i_1\cdots i_{27}}\\
E_7 \subset\Sp(56)& 56 & 133 & A_{ij}, S_{ijkl}, \epsilon_{i_1\cdots i_{56}}\\
E_8 &248 & 248 & \delta_{ij}, A_{ijk},  S_{i_1\cdots i_8}, \epsilon_{i_1\cdots i_{248}}\\
\end{array}$$
\end{center}
\caption{\label{tab:tensori}\em Invariant tensors of Lie groups, where $S$ denotes a symmetric tensor
$A$ denotes an anti-symmetric tensor, and $\epsilon$ is the Levi-Civita tensor.
Complex conjugate representations are shown as upper indices.
We recall that  $\SU(2)=\SO(3)=\Sp(2)$,
$\SO(5)=\Sp(4)$, $\SU(4)=\SO(6)$, $\SO(4)=\SU(2)^2$.
}
\end{table}%

\subsection{Long-lived SO glueballs: group parity}
Longer-lived glueball states arise if the gauge group is $\SO(N)$ with even $N \ge 8$.
Recalling that $\SO(N)$ vectors can be described by an anti-symmetric matrix $\G_{ij} = G^a T^a_{ij}$,
a long-lived special glueball with mass $M \approx N \ldc/2$ is formed by $N/2$ vectors and
corresponds to the gauge-invariant operator 
\beq\label{eq:oddball} {\cal O}_{\rm OB} ={\rm Pf}\, \G  \sim  \epsilon_{i_1\cdots i_N} {\cal G}_{i_1 i_2}\cdots {\cal G}_{i_{N-1} i_N}   \eeq
where Lorentz indices have been omitted, and the Pfaffian is the square root of the determinant.
Such state, being built with the Levi-Civita $\epsilon$ tensor, 
is odd under a $\mathbb{Z}_2$ parity in internal $\SO(N)$ space.
Following~\cite{1907.11228,Buttazzo:2019mvl} we dub it O-parity, in order not to confuse it with usual parity P. 
We dub the O-odd glueballs as `odd-balls'.

O-parity can be concretely represented, for example, as a reflection of the first component of the fundamental representation, such that
it acts on vectors as $\G_{ij}\to(-1)^{\delta_{1i}+\delta_{1j}}\G_{ij}$, leaving
the SO$(N)$ group algebra invariant~\cite{1907.11228,Buttazzo:2019mvl}.
More abstractly, the odd-ball transforms as $ {\rm Pf}\, \G  \to {\rm Pf}\, \G  \times\det R $ under $\SO(N)$ rotations $R$,
where $\det R=\pm 1$ for O$(N)$ groups.
O-parity is an accidental symmetry of the renormalizable gauge action, as well as of its gravitational extension.

\medskip

Under assumptions 1.\ or 2.\ in the introduction, odd-balls are gravitationally stable. 
They decay gravitationally under assumption 3., namely if the action contains the higher-dimensional operator 
${\rm Pf}\,\G/M_{\rm Pl}^{N-4}$, which explicitly breaks O-parity.
Odd-balls decay into SM fields in the presence of operators such as $|H|^2\, {\rm Pf}\,\G/M_{\rm Pl}^{N-2}$.
In both cases, odd-balls decay with rate
\beq\label{eq:GammaOB}
\Gamma_{\rm OB} \sim M  (M/M_{\rm Pl})^{2N-4},\eeq
which is highly suppressed at moderately large $N$.\footnote{A similar highly protected accidental global symmetry
was used to propose a high-quality axion model in~\cite{2007.12663}.}
For example, for $N=10$,
$\SO(10)$ odd-balls can be DM in the mass range $10^{10}\GeV \circa{<} \ldc\circa{<}10^{15}$.
For $N=6$, the long-lived glueball of $\SO(6)=\SU(4)$ is the C-odd glueball of SU(4), and eq.\eq{GammaOB} reduces to eq.\eq{GammaCB}.

\medskip

Odd-balls exist for $\SO(2k)$  groups because the fundamental representation is real, so that vectors can be written as a matrix  ${\cal G}_{ij}$,
with two lower indices, that turns out to be anti-symmetric.\footnote{The ${\cal G}_{ij}$ matrix of a Sp($N$) group
is symmetric so that its contractions with the
$\epsilon$ tensor vanish.  
More in general,  $\epsilon$  is not a fundamental tensor of $\Sp(N)$, as it can be written as
$\epsilon_{i_1\cdots i_{N}} = A_{i_1 i_2}\cdots A_{i_{N-1}i_{N}} + \hbox{permutations}$
in terms of the anti-symmetric invariant tensor $A_{ij}$ with two indices of $\Sp(N)$.}
The list of Lie groups and their invariant tensors in table~\ref{tab:tensori}~\cite{Cvitanovic:1976am,Cvitanovic:2008zz}, 
suggests that similar states exist if $G$ is the exceptional group $F_4$ or $E_8$.
Concerning $F_4$, its ${\cal G}_{ij}$ matrix of vectors is a sub-group of SO(26) vectors with rank 24
(ignoring Lorentz indices and derivatives)~\cite{2011.01764}. 
Concerning $E_8$, its $248\times 248$ matrix of vectors ${\cal G}_{ij}$ has rank 240;
we don't know if the bound state built with the $\epsilon_{i_1\cdots i_{248}}$ tensor
is long-lived or can fragment into smaller bound states
built with the $A_{ijk}$ and $S_{i_1\cdots i_8}$ invariant tensors of $E_8$.

\subsection{Hadronization of vectors}\label{hadron}
In the next section we will consider pairs of vectors  produced by collisions with $E=\sqrt{s}\gg \ldc$.
We here estimate the particle multiplicity in vector jets, determined by 
soft-radiation doubly enhanced by IR logarithmic terms that can be resummed through evolution equations~\cite{hep-ph/9910226,hep-ph/0004215}.
The numerical factor is the same for all groups, as the vector/vector splitting function, proportional to the Casimir $C_G$, gets
cancelled by the one loop running of the gauge coupling in it, with beta function proportional to $C_G$.
As a result
\beq \label{eq:NDG}
N_{\rm DG}(E) \sim \exp\left( \sqrt{\frac{48}{11}\ln \frac{E}{\ldc}}\right).\eeq 
We next need to estimate the relative amount of special glueballs with respect to ordinary glueballs.
It's enough to consider $G=\SO(N)$ and its special odd-balls.
We model hadronization as follows.
We assume that nearby vectors start forming bound states until achieving singlets under $G$.
We perform a MonteCarlo simulation, that adds to a list of vectors one more vector
${\cal G}_{ij}$ with random $i<  j$ indices until singlets are possible.  
Whenever $k>1$ vectors can be contracted with $\delta_{i j}$ tensors, we assume that these $k$ vectors form
a  glueball with $k$ vectors, that drop out of the bound state.  
Whenever the bound state contains $N/2$ vectors with all different indices, we assume that 
these $N/2$ vectors form an odd-ball.
Running numerically up to $N = 16$ we find that the fraction of the dark vector energy that ends up in $\SO(N)$ odd-balls is approximately 
$1.2 \times 0.76^N$, mildly suppressed at large $N$.
The energy fraction that ends up in ordinary glueballs made of $k$ vectors is approximatively
given by the Poisson distribution $e^{-\mu} \mu^{k-1}/(k-1)!$ with $\mu=(N+4)/6$.

\section{DM production from thermal scatterings}\label{thermal}
We here consider DM produced through the freeze-in mechanism by which a hot bath of high temperature SM particles occasionally
results in gravitational collisions that pair-produce dark matter. 
The $2\to2$ annihilation processes SM SM $\to$ DM DM  occur through the $s$-channel exchange of a graviton $h_{\mu\nu}$. Since the production is driven by non-renormalizable gravitational interactions, it is dominated by the highest available energy scales after
the end of inflation, i.e the reheating temperature $\trh$.
Defining the Hubble rate during inflation as $\hi$, the reheating temperature is $\trh\approx (45/4\pi^3 g_{\rm SM})^{1/4}\sqrt{M_{\rm Pl}\hi}$ if SM 
reheating happens instantaneously, or 
smaller $\trh\sim\sqrt{M_{\rm Pl}\Gamma_{\rm infl}}$ otherwise. 
We distinguish two regimes	
\begin{itemize}
\item if $\trh\gg\ldc$ DM is mostly produced in form of  massless dark gluons $\G$ (sections~\ref{gluonsProd}, \ref{reh})
which later undergo cosmological evolution (section~\ref{cosmoVevo}).

\item if $\trh\ll\ldc$ DM is directly produced in form of hadronic bound states; as discussed in section~\ref{sec:DGprod}.
\end{itemize}
In both cases, one needs to take into account the possibility that faster decays of ordinary glueballs reheat the Universe, see section~\ref{dilu}.

\subsection{Thermal production rate of massless dark vectors}\label{gluonsProd}
We expand the metric as $g_{\mu\nu} = \eta_{\mu\nu}  + 2h_{\mu\nu}/\bar M_{\rm Pl}$, where $\bar M_{\rm Pl} = M_{\rm Pl}/\sqrt{8\pi}$
is the reduced Planck mass.
The graviton propagator with quadri-momentum $k$
is $i(\eta_{\mu\mu'} \eta_{\nu\nu'}+\eta_{\mu\nu'}\eta_{\nu\mu'})/2k^2 + \cdots$.
One graviton $h_{\mu\nu}$ couples as
$h_{\mu\nu} T^{\mu\nu} /\bar M_{\rm Pl}$ 
where $T^{\mu\nu} \equiv 2\, {\delta S}/\delta g_{\mu\nu} = T^{\mu\nu}_{\rm SM}+T^{\mu\nu}_{\rm DM} + \cdots$ is the usual energy-momentum tensor.
In our case it is traceless, as we consider massless vectors, and we neglect masses of SM particles
at temperature much above the weak scale.
Tree-level  $s$-channel exchange of one graviton  generates the effective amplitude
$\mathscr{A}= -i {\cal T}/\bar{M}_{\rm Pl}^2 s$ where ${\cal T} = T_{\mu\nu} T^{\mu\nu}$.

The differential cross sections for production of the $d_G $ massless gauge bosons  from scatterings of
particles with negligible masses
and spin $S=\{0, 1/2,1\} $ are obtained as
$\sfrac{d \sigma_S}{dt}=\sfrac{S_f \med{|\mathscr{A}|^2}_S}{16\pi s^2}$,
where $S_f = 1/2$ and
$\med{|\mathscr{A}|^2}$ is summed over the polarizations of the final states and averaged over the initial states.  
Most cross sections can be obtained from~\cite{hep-ph/0408320}\footnote{The cross sections in points 3 and 4  in Appendix A,
there computed for a tower of KK gravitons, are adapted to a single graviton as ${\cal S}(x) = 1/\bar{M}_{\rm Pl}^2 x$.}
or from~\cite{1708.05138} (see also~\cite{Garny:2015sjg}, where some cross sections differ)
\beq
\frac{d \sigma_0}{dt}=
d_G  \frac{t^2 u^2}{16\pi  s^4\bar M_{\rm Pl}^4},\qquad
\frac{d \sigma_{1/2}}{dt}= d_G  \frac{  tu (t^2+u^2)}{64 \pi s^4 \bar M_{\rm Pl}^4},\qquad
\frac{d \sigma_{1}}{dt}=   d_G  \frac{t^4+u^4}{64  \pi s^4 \bar M_{\rm Pl}^4}  
\eeq
where $s =(p_1+p_2)^2$, $t = (p_1 - q_1)^2$, $u =(p_1 - q_2)^2$ are the usual Mandelstam variables.
The non-minimal coupling $\xi$ of the spin-0 scalar $h$ to gravity does not contribute to $d\sigma_0$,
because it adds to the SM energy momentum tensor a term proportional to
$(\partial_\mu \partial_\nu - \eta_{\mu\nu} \partial^2) h^2$ and because
$T^{\mu\nu}_{\rm DM}$ is trace-less and conserved
{(see also~\cite{1410.6436}).}

The interaction rate density in thermal equilibrium at temperature $T$ is 
\begin{equation}\label{eq:gamma}
\gamma_S^{\rm eq}=g_{S}^2S_i \frac{T}{32\pi^4}\int_{s_{\rm min}}^{\infty}ds \, s^{3/2}\sigma_S(s) K_{1}\left(\frac{\sqrt{s}}{T}\right)
\end{equation}
where $S_i=1/2$ if the initial particles are identical (real scalars and gauge bosons) and $S_i=1$ otherwise ({Dirac} fermion/anti-fermion pair); 
$g_{S}$ are the polarization degrees of freedom of the initial particles (1 for scalars, 2 for fermions and massless vectors) 
and $K_{1}$ is the modified Bessel function.
We find
\begin{equation}
\gamma^{\rm eq}_0=\frac{d_G  T^8}{40\pi^5\bar M_{\rm Pl}^4},\qquad
\gamma^{\rm eq}_{1/2}=\frac{3d_G  T^8}{20\pi^5\bar M_{\rm Pl}^4},\qquad\gamma^{\rm eq}_1=\frac{3d_G  T^8}{10\pi^5\bar M_{\rm Pl}^4}.
\end{equation}
The total interaction rate density is
\beq \gamma^{\rm eq}({\rm SM~SM} \to \G\G) =N_0\gamma^{\rm eq}_{0}+N_{1/2}\gamma^{\rm eq}_{1/2}+N_1\gamma^{\rm eq}_{1}
=\frac{{ 283} d_G  T^8}{{ 40}\pi^5\bar M_{\rm Pl}^4} \eeq 
where 
$N_S$ are the number of degrees of freedom, $N_0=4$, $N_{1/2}=45{/2}$ 
and $N_1=12$ in the SM.

\subsection{Freeze-in abundance of massless vectors}\label{reh}
We here compute the freeze-in gravitational production rate of dark vectors with negligible mass, $\ldc\ll T_{\rm RH}$.
Assuming that the big-bang suddenly started at the maximal temperature $T_{\rm RH}$
and that the number abundance of dark vectors $n$ vanishes at $T_{\rm RH}$, its later evolution
is dictated by the Boltzmann equation
\beq  \dot n+ 3 H_R n = 2\gamma^{\rm eq}.\eeq
Here a dot denotes $d/dt$, 
$H_R = \dot a/a = \sqrt{8\pi \rho_R/3}/M_{\rm Pl}$ is the expansion rate,
$\rho_R=\pi^2 g_*  T^4/30$ with $g_*=106.75$
is the energy density of SM radiation at temperature $T$.
It is convenient to rewrite the Boltzmann equation in terms of $Y = n/s$
as function of $z= T_{\rm RH}/T$,
where $s=4\rho_R/3T$ is the SM entropy density.
The resulting Boltzmann equation is $sH_Rz \, dY/dz=2\gamma^{\rm eq}$, solved by
\beq Y(T\ll T_{\rm RH}) = \left.
\frac{2\gamma^{\rm eq}}{3H_Rs}\right|_{T=T_{\rm RH}} .\eeq 
A similar result is found with a more realistic definition of $T_{\rm RH}$,
that assumes that SM particles are progressively reheated by the energy
released by some non-relativistic energy density $\rho_\phi$
that decays with width $\Gamma_\phi$ into SM particles only.
$\rho_\phi$ could be due to the inflaton, or to some non-relativistic unstable particle.
The Boltzmann equations now are
\beq\label{eq:Boltzt}
\dot\rho_\phi + 3H\rho_\phi = -\Gamma_\phi \rho_\phi ,\qquad
\dot\rho_R + 4 H \rho_R = \Gamma_\phi \rho_\phi ,\qquad
 \dot n+ 3 H n = 2\gamma^{\rm eq}
\eeq
where $H =  \sqrt{8\pi(\rho_\phi  + \rho_R)/3}/M_{\rm Pl}$. 
The reheating temperature $T_{\rm RH}$ is defined in terms of $\Gamma_\phi$ as the temperature at which
\beq
\Gamma_\phi = H_R(T_{\rm RH})
\qquad\hbox{i.e.}\qquad
T_{\rm RH}= \left[ \frac{45}{4\pi^3 g_*}\,\Gamma_\phi^2
M_{\rm Pl}^2\right]^{1/4}  .\eeq
What happens at $T\gg T_{\rm RH}$ gets diluted by the entropy release.
Numerical solutions give 
\beq \label{eq:Yres} Y(T\ll T_{\rm RH}) = 0.53 \left.
\frac{2\gamma^{\rm eq}}{3H_Rs}\right|_{T=T_{\rm RH}} = 1.2~10^{-22}
\left(\frac{T_{\rm RH}}{10^{12}\GeV} \right)^3
\frac{2\gamma^{\rm eq}|_{T=T_{\rm RH}}}{T_{\rm RH}^8/\bar M_{\rm Pl}^4}
\approx { 7.4~10^{-5}} d_G  \left( \frac{T_{\rm RH}}{\bar M_{\rm Pl}}\right)^3. \eeq 

 \begin{figure}[t]
 \centering
 $$\includegraphics[width=0.8\textwidth]{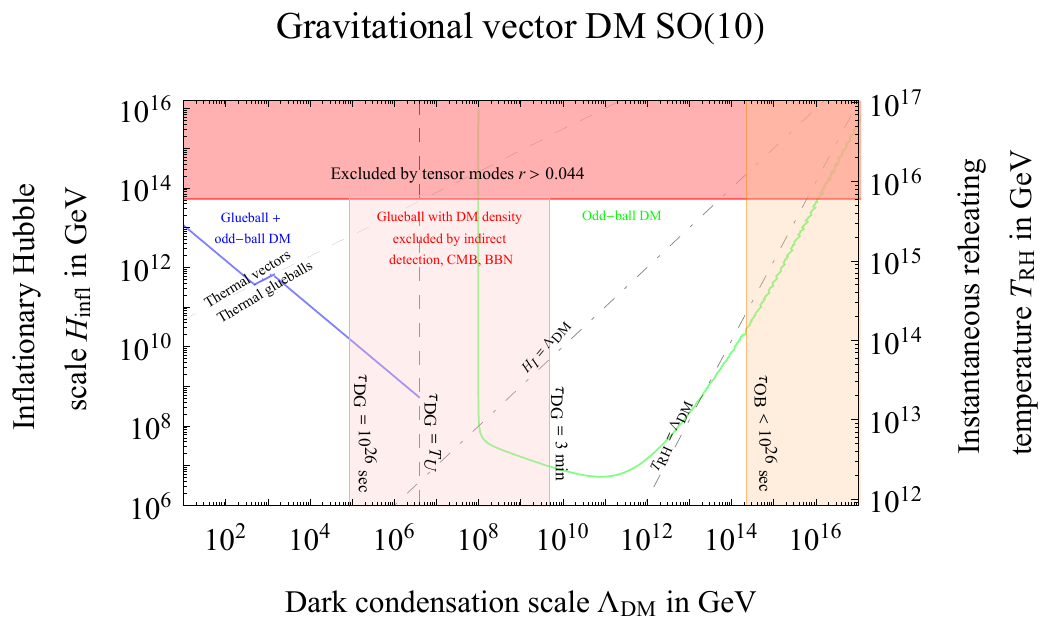}$$
 \vspace{-1cm}
 \caption{\em\label{fig:plot1}Parameter space of the $\SO(10)$ model. The DM abundance is reproduced along the blue curve (in which case it is composed of both ordinary dark glueballs and odd-balls) 
 and along the green curve (in which case it is composed of odd-balls only, since other glueballs have decayed). 
The upper red region is excluded from the current bound on the tensor inhomogeneities. 
Finally, assuming generic Planck-suppressed operators in the pink and orange regions 
some state decays with lifetime between $3\,{\rm min}$ and $10^{26}\,{\rm sec}$:
these regions are excluded by indirect detection, CMB and Big Bang Nucleosynthesis constraints
if the decaying state has a relic abundance comparable to DM.}
 \end{figure}

\subsection{Cosmological evolution of dark vectors}\label{cosmoVevo}
After vectors are gravitationally pair-produced with  small number density $Y=n/s \sim (T_{\rm RH} /\bp)^3$ and energy density $\rho \sim T_{\rm RH} n $,
three qualitatively different cosmological evolutions can take place, depending on the value of $\ldc$:
\begin{itemize}
\item For large $\ldc$ (we determine below how large)
vectors don't thermalize and their number or energy density are too small to thermally block confinement.
Thereby pair-produced vectors immediately hadronise: each vector forms 
$N_{\rm DG}(T_{\rm RH})$ dark glueballs with mass $\sim \ldc$,
with  $N_{\rm DG}$ estimated previously in eq.\eq{NDG}.
This happens when the number density  $ \sim N_{\rm DG}  n $ is smaller than $\ldc^3$, and
thereby for large $\ldc\circa{>} N_{\rm DG}^{1/3} (\trh^2/\bp)$. 
In this regime, glueball self-interactions are slower than the Hubble rate;
ordinary glueballs next decay quickly, giving negligible cosmological effects. 
Longer-lived glueballs remain as thermal relics; their density gets diluted by universe expansion while their $Y=n/s$ remains constant.
\end{itemize}
For smaller $\ldc$ the vector energy density is large enough to prevent immediate hadronization:
\begin{itemize}
\item  For very low $\ldc \circa{<}  \bp(\trh/\bp)^{15/4}$ screening of color charges blocks confinement 
and reduces initial-state radiation, that we neglect: $N_{\rm DG}\sim 1$.
Vector thermalization takes place when the SM sector has temperature $T^{\rm therm}_{\rm SM}\sim \bp(\trh/\bp)^3$, because
at this point the scattering rate among vectors (with cross sections $\sigma\sim  \gdc^4/T^2$)
exceeds the Hubble rate.
Energy conservation implies that vectors thermalize to temperature $T_{\rm DM}^{\rm therm} \sim  \bp (\trh/\bp)^{15/4}$.
Next vectors cool and hadronise when their temperature falls below $T_{\rm DM}\sim\ldc$,
acquiring thermal abundance $Y_{\rm DM}\sim (T_{\rm DM}/T_{\rm SM})^3\sim (\trh/\bp)^{9/4}$. 
We can neglect cannibalistic $3\leftrightarrow 2$ processes that occur after hadronization, as they only give a logarithmic correction to $Y_{\rm DM}$.

\item The intermediate range $\bp(\trh/\bp)^{15/4}\circa{<}\ldc\circa{<}\trh^2/\bp$ is similar to the previous case,
except that vectors start interacting through non-perturbative interactions, that lead to their confinement when
the universe expanded enough that $n \sim \ldc^3$,
corresponding to a temperature of the SM sector equal to $T^{\rm had}_{\rm SM}  \sim \ldc(\bp/\trh)$.
At this point vectors form glueballs with mass $\sim\ldc$ and thermal energy $\sim T^{\rm had}_{\rm DM}$.  
Their abundance gets later corrected by
cannibalistic processes, that we can again neglect.
\end{itemize}

\subsection{Thermal production and decay of dark bound states}\label{sec:DGprod}
Finally, we consider thermal production at $\trh \circa{<}\ldc$ of dark bound states. 
As these are unstable, the production rate via inverse decays is simply given in terms of their decay width as
\begin{equation}\label{eq:DGprod}
\gamma_{\rm eq} ({\rm SM~SM} \to {\rm DG})=\frac{T \mdg ^2}{2\pi^2} K_1\left(\frac{\mdg }{T}\right)\GammaDG.
\end{equation}
Model-independent bounds imply that DM freeze-in due to inverse decays negligibly contributes to the DM density. 
Therefore the production of DM for temperatures $T\circa{<}\ldc$ is dominated by the exponentially suppressed tail of
production of vectors.

\subsection{Glueball decays and DM dilution}\label{dilu}
If ordinary dark glueballs have lifetime long enough that they dominate the energy density of the Universe while decaying into SM particles and gravitons, they substantially reheat the Universe and dilute the abundance of stabler DM odd-balls.
The dilution effect is sizeable 
if $\ldc\ll\trh^2/M_{\rm Pl}$ and $\tauDG<T_{\rm U}\sim10^{10}\,{\rm yr}$. In this region glueballs decay at temperature  
\begin{equation}
T_{\rm decay} \approx \frac{\ldc^3}{\bp^2}
\begin{cases}
(\bp/\trh)/N_{\rm DG}^{1/3} \qquad   \text{if glueballs are not thermal}\\
(\bp/\trh)^{3/4} \qquad \ \ \ \ \text{if glueballs are thermal} \,. \\
\end{cases}
\end{equation}
We can approximate decays as instantaneous, finding  that they
reheat the Universe up to $T_{\rm RH}'\sim \ldc\left(\sfrac{\ldc}{M_{\rm Pl}}\right)^{3/2}$. 
Then, the DM abundance gets diluted down to $Y_{\rm DM}^{\rm diluted}=Y_{\rm DM} D_{\rm DG}$ where the dilution factor $D_{\rm DG}$
is estimated as
\begin{equation}
D_{\rm DG}\sim
\begin{cases}\displaystyle
\frac{1}{ N_{\rm DG}}\left(\frac{\sqrt{\ldc \bp}}{\trh}\right)^3 \qquad&  \text{if glueballs are not thermal,}\\ \displaystyle
\left(\frac{{\ldc^2 \bp}}{\trh^3}\right)^{3/4} \qquad & \text{if glueballs are thermal.}\\
\end{cases}
\end{equation}
Combining all effects above, and considering a model with $G=\SO(10)$, gives the numerical result in fig.\fig{plot1},
that shows that gravitational freeze-in can reproduce the DM abundance in two different regimes:
either as ordinary glueballs lighter than about 100 TeV (see also~\cite{2011.10565}), or as special long-lived odd-balls, that can thereby be much heavier,
up to $10^{14}\GeV$ in the figure.
The intermediate mass range around $10^7\GeV$ is excluded by bounds on the decaying ordinary glueballs,
if the DM abundance is achieved.
In this intermediate range dilution is sizeable and the DM abundance is proportional to $Y_{\rm DM}/Y_{\rm DG}$ where $Y_{\rm DG}$ is the abundance of the decaying ordinary glueballs. This ratio does not depend on $\trh$, so that requiring that the measured DM abundance is reproduced 
fixes the value of $\ldc$ ($\ldc\approx10^8\GeV$ for $N=10$).

\medskip

In the next section we will show that pure gravitational production during inflation gives a negligible extra contribution to the DM density.

\section{DM production from inflationary fluctuations}\label{inflation}
Freeze-in gravitational production during the Big Bang is dominated by the highest temperatures around $T_{\rm RH}$, which  means that earlier cosmology can contribute more if higher energies were present.
We assume an earlier inflation phase, during which
the energy density of the Universe was dominated by a scalar field~$\phi$, the inflaton.
Its energy density is $\rho = \dot \phi^2/2 + V$ and its pressure is $p =\dot \phi^2/2 -V$.
The inflaton potential $V(\phi)$ allows for a prolonged rapid expansion (usually on a plateau, the so-called slow-roll phase), followed by an oscillatory phase which sets the reheating temperature $T_{\rm RH}$, where the inflaton acquires its mass $m_\phi$.
DM production can arise in multiple ways:
\begin{enumerate}[a)]
\item from quantum fluctuations during inflation;
 
\item after inflation end, when the inflaton is possibly oscillating around the minimum of its potential.
In our context the inflaton couples to DM at least gravitationally, and the phase space is open if
$m_\phi \circa{>} \hbox{${\cal O}$(1)}\times\ldc$;

\item when the inflaton finally decays, if it decays into DM. 
In our context this happens at least gravitationally if $m_\phi \circa{>} 2\hbox{${\cal O}$(1)}\times\ldc$.
\end{enumerate}
In the above list, b) and c) are particle physics processes (see~\cite{1804.07471}) that depend on the inflaton model.
In particular, if the inflaton decays gravitationally into both SM and DM, 
c) would overwhelm the scattering contribution computed in the previous section.
We assume instead that $m_\phi$ is small enough that b) and c) can be neglected,
and focus instead on the purely inflationary production a).

During inflation, the negative pressure of the inflation (thanks to its barotropic parameter $w \approx -1$) drives the exponential expansion of the Universe as dark energy with scale factor $a \propto e^{H_{\rm infl} t}$ with a roughly constant 
$H_{\rm infl}  = (8\pi V/3M_{\rm Pl}^2)^{1/2}$.
The Universe, even if somewhat inhomogeneous at small scales, quickly ends up being described by the FLRW metric 
\begin{align}
ds^2 = g_{\mu\nu}dx^\mu dx^\nu  = dt^2 - a(t)^2  dx^2.
\end{align}
As such, we expect inflation to homogenize the field relatively quickly.

 \begin{figure}[t]
 \centering
 $$\includegraphics[width=0.65\textwidth]{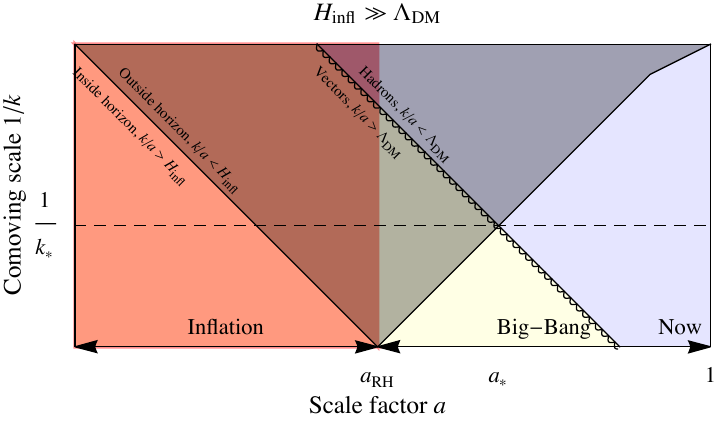}$$
 \vspace{-1cm}
 \caption{\em\label{fig:InflationScales} Evolution during and after inflation of the relevant scales,
 assuming instantaneous reheating at inflation end, when $a = a_{\rm RH}$.
The black continuous curve is the horizon, and the  region beyond the horizon is shaded in gray.
The half-wavy line is the transition from vector to hadron modes.
The special mode $k_*$ is the one that, after inflation at $a = a_*$, 
re-enters the horizon while transitioning from vector to hadron.}
 \end{figure}

The FLRW spacetime is conformally flat, so that gravity does not couple to massless classical vectors  that enjoy a conformal symmetry.
At quantum level, conformal invariance is broken by the RG running of $\gdc$, which generates
a non-vanishing trace for the energy-momentum tensor,
$T_{\mu}^{\mu} =  \beta_{g_{\rm DM}} \Tr \,\G_{\mu\nu}^2/2\gdc$, and ultimately leads to confinement
and dynamical generation of the mass scale $\ldc$.
However, we cannot perform lattice simulations of strongly-interacting vectors during Universe expansion,
and inflationary production is usually computed in terms of free modes.
We thereby approximate the system by making explicit its quasi-free degrees of freedom in the following unusual effective action 
\beq 
\label{eq:effact1}
S^{\rm trans}_{\rm eff} = \int d^4x\,\sqrt{-\det g}
\left\{\begin{array}{ll}
-\sfrac{\hat G^{a \mu\nu} \hat G_{a \mu\nu}}{4\gdc ^2} & \hbox{for modes with $k/a\circa{>}\ldc$}\\
\sum_X [(\partial X)^2 - (m_X^2+\xi_X R) X^2]/2  & \hbox{for modes with $k/a\circa{<}\ldc$}\\
\end{array}\right.\eeq
where $X$ are the multiple quasi-stable dark hadrons with masses  $m \sim\ldc$ and unspecified spins $0,1,\ldots$.
Eq.\eq{effact1} means that, when expanding fields into modes, we only retain hadrons at low energy and vectors at high-energy;
for these we approximate the scale anomaly by using the running gauge coupling $\gdc$ renormalized at their energy scale.
As usual when computing in cosmology, we define $k$ as the comoving momentum, so that $k/a$ is the momentum.
If $H_{\rm infl} \circa{<}\ldc$ the vector description is never relevant (modes `confine' early during inflation),
and the computation reduces to inflationary production of massive dark hadrons $X$, to be studied in section~\ref{lowH}.
If $H_{\rm infl} \circa{>}\ldc$ modes cross inflation end while being in the vector description
and transition to the hadron description later: 
fig.\fig{InflationScales} sketches how the relevant scales evolve,
and the computation is performed in section~\ref{hiH}.\footnote{Production of vector dark matter during inflation was studied in~\cite{Ahmed:2020fhc}, considering
an Abelian massive vector with Stueckelberg mass: in this case inflation dominantly produces its longitudinal component, 
which can even have a ghost-like behaviour.
We consider a qualitatively different situation: non-abelian massless confining vectors with two transverse components.}

\subsection{Inflationary production at $H \circa{<} \ldc$}\label{lowH}
In this regime inflation can produce dark hadrons $X$.
We focus on a massive scalar hadron bound state, as massive higher-spin bound states only exist in the low-energy regime where
the peculiar behaviour of massive fundamental higher-spin states (e.g.\ longitudinal modes of spin~1, see~\cite{1504.02102,1812.00211,1905.09836,2009.03828}) does not take place.
Then the bound-state field equations in the homogenous limit are
\begin{align} 
\ddot X + 3H \dot X + (m^2+\xi R) X = 0
\end{align}
where $R=12 H_{\rm infl}^2$ during inflation. The non-minimal coupling to gravity $\xi$ is, in principle, computable~\cite{Hill:1991jc}, { and depending on the details of the model, can also lead to production of DM during the preheating stage~\cite{Karam:2020rpa}.}
Non-perturbative dynamics is not expected to produce the special conformal value $\xi =1/6$,
as strong coupling breaks the conformal symmetry.
In order to examine how dark matter is produced, we perturb the field around the homogeneous value $\langle X\rangle=0$ as~\cite{1903.08842}
\beq X(\vec x, t) = \int \frac{d^3k}{(2\pi)^3} \frac{e^{i \vec k\cdot \vec x}}{a^{3/2}} \bigg[X_k (t) a_{\vec k} + X_k^*(t) a_{-\vec k}^\dagger \bigg].\eeq
Neglecting time derivatives of $\hi$,
the equations of motion for the perturbation components $ X_k$ with comoving momentum $k$ are
\begin{align}\label{eq:Xkevo}
 \ddot X_k  + \omega_k^2  X_k = 0,\qquad 
\omega_k^2 = \frac{k^2}{a^2} + \hi^2 \mu^2,\qquad
\mu ^2\equiv -\frac14+ \frac{m^2}{H^2_{\rm infl}}  + 12 \left(\xi-\frac16\right) .
\end{align}
The equations for $X_k$ are solved in terms of Bessel functions of order $i\mu$.
Particle production during inflation can be computed at the end of inflation by using the in-out formalism: there are two classes of solution to the equation of motion, $X^{\rm in}_k$ and $X^{\rm out}_k$, found by imposing as a boundary condition
the Bunch--Davies vacuum  at very early times $aH \ll k$
\begin{align} 
(X_{k})_{\rm BD} \simeq \frac{\exp(ik/aH)}{\sqrt{2k}} .
\end{align}
Since $a$ depends on time, the vacuum is time-dependent, and so the mode functions evolve according to the Bogoliubov transformation
\begin{align} 
X^{\rm in}_k = \alpha_k X^{\rm out}_k + \beta_k  X^{\rm out}_k.
\end{align}
The density of the final products is given by $\bra 0_{\rm in}| a^\dagger_{\rm out}({\vec p}) a_{\rm out}({\vec q})|0_{\rm in} \ket = |\beta_k|^2 \delta^{(3)}({\vec p}-{\vec q})$, which  means
\begin{align}\label{nX}
N_X = \int_0^{\infty} dk \, 2\pi k^2 |\beta_k|^2.
\end{align}
The Bogoliubov factor that relates in and out states is
$
|\beta_k|^2 =  1/(e^{2\pi\mu}-1)
$.
The integral in eq. \eqref{nX} is divergent. This is to be expected, since it corresponds to the total density of particles produced across all time.  Moreover, the bulk of particle production occurs when the two terms in $\omega_k$ are equal, corresponding to a vector mode with a wavelength set by its effective mass. We may then use the time at which production reaches its peak for each mode in order to write the integral for $N_X$ in terms of $\eta$. This is given by $k/(a\hi) = \mu$, or at conformal time $\eta = - \mu/k$, and so
\begin{align} 
N_X &= 2\pi |\beta_{k}|^2 \mu^3   \int^0_{-\infty} d\eta \,(aH)^4.
\end{align}
The density of the ``out'' states at the end of inflation will then corresponds to the physical DM density. We are interested in the production rate per unit (physical) volume $\Gamma$, which is related to the physical number density $n_X$ at any given moment by $n_X = \Gamma/(3H)$.
Using the expression of the Bogoliubov coefficient $\beta_k$, we can read off the rate $\Gamma$, which 
leads to the result that the density of bound states at inflation end roughly corresponds to a thermal density with $T \sim H_{\rm infl}/2\pi$~\cite{hep-ph/0104100,1804.07471,1903.08842,2009.03828},   
\beq n_{\rm infl}= \frac{\hi^3}{3} \frac{2\pi \mu^3}{e^{2\pi\mu}-1} \sim  H^3_{\rm infl}   e^{-m/H_{\rm infl}},\qquad
\rho_{\rm infl} \approx m n_{\rm infl}.\eeq
Let us discuss the two limits of this result.
\begin{itemize}
\item Production of heavy bound states with masses $m \gg \hi$ is exponentially suppressed as expected.
Assuming $m\sim\ldc$ and instantaneous reheating (which then initiates radiation domination) with $T_{\rm RH}\sim \sqrt{M_{\rm Pl} \hi}
\gg \hi$ implies that the inflationary contribution is sub-dominant with respect to the contribution 
of freeze-in gravitational collisions,
\beq  \label{eq:ratiocollinfl}
\frac{\rho_{\rm infl}}{\rho_{\rm coll}} \sim \sqrt{\frac{\hi}{M_{\rm Pl}}} e^{-\ldc /\hi} \ll1 .\eeq
\item Production of scalars so light that $\mu^2<-1/4$ is not covered by this computation.
Indeed, $\mu^2 = -1/4 +\sfrac{m_{\rm eff}^2}{\hi^2}$ where
$m_{\rm eff}$ is the effective mass of $aX$ as function of conformal time:
this is the Weyl transformation that shows that a scalar with $m_{\rm eff}=0$ i.e.\ $\mu^2 = -1/4$  remains in its vacuum state.
\end{itemize}
Fundamental light scalars with $\mu^2 < -1/4$
would develop a vacuum expectation value, giving rise to the `misalignment mechanism' for DM production.
Our scalars $X$ are however hadrons, and,
while we cannot compute what happens if $\ldc \circa{<} \hi$, 
 the deep regime $\ldc \ll \hi $ can be computed by switching to the constituent vector description.
This is studied in the next section.

\subsection{Inflationary production at $H \gg \ldc$}\label{hiH}
In this regime relevant modes exit from inflation while still being described by perturbative vectors, rather than by hadrons.
The vector action that effectively accounts for the quantum RGE running is eq.\eq{effact1}. 
The vector equations of motion  (we can drop non-abelian terms) are
\begin{equation}
\nabla_\sigma \hat G^{\sigma\rho} + \frac{\nabla_\sigma (\gdc ^2)}{\gdc ^2} \hat G^{\sigma\rho} = 0.
\end{equation} 
The coupling $\gdc$ runs with the length scale as
\begin{equation}\label{rungDC}
\epsilon\equiv  \frac{d  \ln \gdc^2}{d\ln a} =
   \frac{11C_G}{3}\frac{ \gdc^2}{8\pi^2}  > 0
\end{equation} 
where $d\ln a= H dt= - dN$ corresponds to the number of $e$-foldings.
We thereby have $\epsilon \ll 1$ in the perturbative regime, and $\epsilon \sim 1$ around confinement.
As expected, the trace anomaly disappears when the running of $\gdc$ is neglected.
Eliminating the temporal component $\hat G_0$ along with the divergence of the spatial components $\hat{G}_i$,
the classical equation for the homogeneous transverse spatial components of vectors, $\hat G_T$, is
\begin{equation}
\frac{d^2 \hat G_T }{dt^2}+ (1-\epsilon)  \hi  \frac{d\hat G_T}{dt} = 0.
\end{equation}
We expand the transverse $\hat{G}_T$ in terms of canonically normalized modes as
\beq \hat G_T(\vec x, t) =   \int \frac{d^3k}{(2\pi)^3}  \frac{e^{i \vec k\cdot \vec x}}{a^{1/2}} \bigg[G_k (t) b_{\vec k} + G_k^*(t) b_{-\vec k}^\dagger \bigg]\gdc.\eeq
Neglecting time derivatives of $\hi$,
the equations of motion for the perturbation components~$G_k$ with comoving momentum $k$ are
\begin{align}\label{eq:Gkevo}
  \ddot G_k  + \omega_k^2  G_k = 0,\qquad \omega_k^2 = \frac{k^2}{a^2} + \hi^2 \mu^2,\qquad
\mu ^2\equiv -\frac14 + \frac{\epsilon}{2} 
\end{align}
where the value of $\epsilon$ is evaluated around horizon exit, so that eq.\eq{RGEsol} gives
$\epsilon \approx 1/\ln(a_*/a_{\rm RH})$
in the notations of fig.\fig{InflationScales}. 
Eq.\eq{Gkevo} has the same form as  eq.\eq{Xkevo} for scalar fluctuations $X_k$, and 
is thereby similarly solved
by Bessel functions of order $i\mu$.
Indeed, in the limit $\epsilon=0$,  eq.\eq{Gkevo} reduces to $\mu^2=-1/4$, the value 
that corresponds to a conformally coupled state
that remains in its ground state despite Universe expansion.
The $\epsilon$ term coming from the RG running of~$\gdc$ makes $\mu^2 > -1/4$, bringing the vector into the regime where it does
not develop a coherent vacuum expectation,
and its inflationary abundance is suppressed by a multiplicative factor of~$\epsilon$ (see eq.~(29) in~\cite{Dimopoulos:2006ms}).

As $\epsilon$ increases, the vector abundance remains small
($|\beta_k|^2 \sim  \sfrac{1}{(2 \pi  \mu)}$ and $n_{\rm infl} \sim  \mu^2 \hi^3$) 
as long as $\epsilon$ barely exceeds $1/2$, such that $\mu$ is small.
Only in the deep non-perturbative regime $\epsilon \approx 1$ does the vector abundance become comparable to the bound-state abundance
computed in section~\ref{lowH}. The two regimes therefore smoothly connect.

\medskip

In conclusion, the above demonstrates that the purely inflationary production of both free vectors and bound states, roughly given by a
thermal density with $ T \sim\hi$ or less,
is negligible compared to freeze-in due to gravitational scatterings,
\begin{align} 
\frac{\rho_{\rm infl}}{\rho_{\rm coll}} \circa{<}  \mu^2 \sqrt{\frac{\hi}{M_{\rm Pl}}}  \ll1 .
\end{align} 
As a result, the final result remains as shown in fig.\fig{plot1}, where only gravitational freeze-in is included.
 
 \section{Conclusions}\label{concl}
We extended the Standard Model  by adding a new `dark' non-abelian gauge interaction and nothing else, 
in particular no matter fields charged under the new interaction.
Then the new dark sector has no renormalizable interactions to the SM and confines at a scale $\ldc$.
We thereby obtained a dark sector that interacts with the SM sector
only gravitationally.
We also allowed for Planck-suppressed non-renormalizable operators.

The dark sector forms various glueball bound states that can be long-lived enough to be Dark Matter candidates.
The simplest glueballs (such as the lightest state $\Tr \,[{\cal G}_{\mu\nu}{\cal G}^{\mu\nu}]$ with $J^{PC} = 0^{++}$)
decay gravitationally  with  lifetime $\tau \sim M_{\rm Pl}^4/\Lambda^5_{\rm DM}$ and can be DM candidates only if $\Lambda_{\rm DM} $ is below 100 TeV.

We have shown that other special glueball states are odd under accidental symmetries of group-theoretical nature.
Such special states are exactly stable in the limit where Planck-suppressed operators are neglected, and otherwise have Planck-enhanced lifetimes.
$\SU(N)$ gauge theories contain special states odd under group charge-conjugation
with $\tau \sim M_{\rm Pl}^8/\Lambda^9_{\rm DM}$.
Longer lifetimes $\tau=  (M_{\rm Pl}/\ldc)^{2N-4}/\ldc$ are obtained in $\SO(N)$ theories at $N \circa{>}8$
because some ``odd-ball'' states are odd under parity in group space.
Similar states might exist for the $F_4$ and $E_8$ exceptional groups.

We compute the relic abundance generated by gravitational freeze-in and by purely inflationary fluctuations,
assuming that model-dependent decays and scatterings of the inflaton into DM are negligible (for example because blocked by a too small inflaton mass).
In particular, inflationary production of non-abelian vectors is possible because quantum corrections break
the conformal symmetry of the classical action, necessitating an unusual computation.
We found that gravitational freeze-in dominates for instantaneous reheating, and that a theory of
non-abelian dark vectors provides successful gravitational DM candidates, either as
ordinary glueballs for $\ldc\circa{<}100\TeV$, or as special heavier glueballs,
For example, fig.\fig{plot1} considers a SO(10) theory, showing that odd-ball DM is allowed for
$10^{10}\GeV \circa{<}\ldc\circa{<} 10^{15}\GeV$.

\medskip

{
Concerning phenomenology, DM signals generically get suppressed down to unobservable levels
when DM becomes heavy and weakly coupled.
Although our model is gravitationally coupled, the weak coupling can be circumvented in two situations.
First, slow gravitational decays of ultra-heavy relics can give signals in indirect detection experiments.
When a ultra-heavy particle decays into some SM particle, electro-weak and QCD
log-enhanced quantum corrections generate a shower containing all SM particles.\footnote{
The neutrino flux from decays of ultra-heavy relics cannot explain the ANITA anomalous events, see~\cite{1904.13396}.
}
Second, two quasi-stable odd-balls can scatter with cross section $\sigma \sim 1/\ldc^2$ producing glue-balls that 
quickly decay gravitationally;
the resulting flux of ultra-energetic SM particles in the Milky Way 
(size $R \sim 10\,{\rm kpc}$, DM density $\rho\sim \GeV/\cm^3$)
is however small because DM is heavy:
$\Phi \sim R_\odot (\rho_\odot/\ldc)^2 \sigma \sim (10^{10}\GeV/\ldc)^4 10^{-18}/{\rm km}^2\,{\rm yr}$.
Furthermore, sub-leading inflationary production can give rise to iso-curvature inhomogeneities.
These effects can easily be much below current bounds}.

\smallskip

To conclude with a curiosity, we give a positive answer to the question raised by Dyson~\cite{DYSON:2013jra}:
no fundamental reason prevents observing a graviton, and thereby confirming its quantum nature.
Indeed, ultra-heavy DM particles could decay into high-energy gravitons, possibly giving
a graviton flux as high as allowed by bounds on extra relativistic radiation. 
This is detectable in principle (although planet-scale detectors are needed)
by gravitational scatterings on matter with cross section $\sigma \sim e^2/M_{\rm Pl}^2$.

\small

\paragraph{Acknowledgements}
We thank G. Corcella, R. Fonseca, Hong-Jian He,
L. di Luzio, M. Redi, F. Sala, M. Strassler and D. Teresi for discussions.
This work was supported by the ERC grant 669668 NEO-NAT and by PRIN 2017FMJFMW.

\end{document}